\documentclass[12pt]{article}
\usepackage{amsmath,amssymb,epsfig,color}
\usepackage{feynmf}
\usepackage{feynmf}
\usepackage{psfrag}
\usepackage{graphicx}
\usepackage{bm}

\newcommand{\be}{\begin{equation}}  
\newcommand{\ee}{\end{equation}} 
\def\slash#1{#1\!\!\!/\!\,\,}  
\newcommand{\nl}{\nonumber \\ }

\newcommand{\order}{{\cal O}}
\long\def\symbolfootnote[#1]#2{\begingroup%
\def\thefootnote{\fnsymbol{footnote}}\footnote[#1]{#2}\endgroup}

\allowdisplaybreaks

\begin{document}

\begin{fmffile}{feynmffile} 
\fmfcmd{%
vardef middir(expr p,ang) = dir(angle direction length(p)/2 of p + ang) enddef;
style_def arrow_left expr p = shrink(.7); cfill(arrow p shifted(4thick*middir(p,90))); endshrink enddef;
style_def arrow_left_more expr p = shrink(.7); cfill(arrow p shifted(6thick*middir(p,90))); endshrink enddef;
style_def arrow_right expr p = shrink(.7); cfill(arrow p shifted(4thick*middir(p,-90))); endshrink enddef;}

\begin{titlepage}

\begin{flushright}
EFI Preprint 11-28\\
October 31, 2011
\end{flushright}

\vspace{0.7cm}
\begin{center}
\Large\bf 
Universal behavior in the scattering of heavy, weakly interacting dark matter on nuclear targets 
\end{center}

\vspace{0.8cm}
\begin{center}
{\sc   Richard J. Hill\symbolfootnote[1]{richardhill@uchicago.edu} and Mikhail P. Solon\symbolfootnote[2]{mpsolon@uchicago.edu}}\\
\vspace{0.4cm}
{\it Enrico Fermi Institute and Department of Physics \\
The University of Chicago, Chicago, Illinois, 60637, USA
}
\end{center}
\vspace{1.0cm}
\begin{abstract}
  \vspace{0.2cm}
  \noindent
  Particles that are heavy compared to the electroweak scale ($M \gg m_W$), and that are
  charged under electroweak $SU(2)$ gauge interactions display universal properties 
  such as a characteristic fine structure in the mass spectrum induced by electroweak symmetry breaking, 
  and an approximately universal cross section for scattering on nuclear targets.  
  The heavy particle effective theory framework is developed to compute these properties.  As illustration, 
  the spin independent cross section for low-velocity scattering on a nucleon is evaluated  
  in the limit $M \gg m_W$, including complete leading-order matching onto quark and gluon operators, 
  renormalization analysis, and systematic treatment of perturbative and hadronic-input uncertainties. 
\end{abstract}
\vfil

\end{titlepage}

\section{Introduction}

Cosmological evidence for dark matter consistent with thermal relic 
Weakly Interacting Massive Particles (WIMPs) motivates laboratory searches 
for such particles interacting with nuclear targets.   
Search strategies and detection potential are highly dependent on the WIMP
mass, $M$,  and its interaction strength with nuclear matter. 
We consider here the class of models where the  
WIMP belongs to an electroweak $SU(2)$ multiplet.
This study is motivated in part by the observation that an exact discrete 
parity arises in Standard Model  
extensions involving confined fermions coupled to electroweak $SU(2)$~\cite{Bai:2010qg}.    
The parity ensures stability of the lightest pseudo-Nambu Goldstone 
mode, which is the electrically neutral component of a Lorentz scalar, electroweak $SU(2)$ 
isotriplet~\cite{DMmodels}. 

Regardless of the origin for such an $SU(2)$ multiplet, e.g. whether it is a composite or
fundamental particle, universal behavior emerges in the limit where the WIMP mass is large 
compared to the electroweak scale, $M \gg m_W$.   
The emergence of these universal properties, and corrections to them, can be 
systematically analyzed using techniques
of heavy particle effective theories~\cite{Isgur:1989vq}.  
We focus on the case of a real scalar transforming as a triplet of electroweak $SU(2)$, 
although the results extend straightforwardly to arbitrary $SU(2)$ representations, 
and to higher spin particles.

At energy scales large compared to $m_W$, the new particle is described
by an effective heavy particle $SU(2)$ gauge theory, 
\be
{\cal L}_{\rm eff} = \phi_v^* ( i v\cdot D + \dots ) \phi_v \,,
\ee
where $\phi_v$ is a scalar heavy particle field, $v^\mu$ is the heavy particle velocity, 
and $D_\mu$ is the $SU(2)$ covariant derivative.
The leading interactions are thus universal, and corrections depending on 
the mass, or other characteristics of the dark matter particle, are suppressed by powers 
of $M$.   In this paper we determine the general 
structure of the heavy scalar effective theory through $\order(1/M^3)$.
As an illustrative application, we compute the
universal cross section for low-velocity scattering of $SU(2)$-charged WIMPs on a nucleon
in the limit $M \gg m_W$.   We present a complete leading order matching onto gluonic operators, 
renormalization analysis, and systematic treatment of perturbative and 
hadronic-input uncertainties. 

The remainder of the paper is structured as follows.
In Section~\ref{sec:hpeft}, we construct the relevant heavy particle effective
theory at scales $\mu \gg m_W$, and compute the leading Wilson coefficients 
in a simple model.  
In Section~\ref{sec:loweft} we consider the operator basis and mass corrections in 
the low energy theory after integrating out scales $\mu \sim m_W$.  
In Section~\ref{sec:match} we perform explicit matching computations at the scale $\mu \sim m_W$. 
In Section~\ref{sec:RG} we perform renormalization group evolution from $\mu\sim m_W$ 
down to low scales $\mu \approx 1\, {\rm GeV} \gtrsim \Lambda_{\rm QCD}$, 
where QCD matrix elements are estimated. 
Section~\ref{sec:xs} presents the cross section for low-velocity scattering
on a nucleon.  Section~\ref{sec:summary} presents a summary and outlook. 

\section{Heavy scalar effective theory: electroweak symmetric theory \label{sec:hpeft} }

Consider the effective theory and matching conditions 
for a scalar particle of mass $M$, charged under electroweak $SU(2)$. 
With obvious modifications, the construction applies to general gauge groups. 
We start by investigating the effective theory at scales $m_W \ll \mu \ll M$, 
with unbroken electroweak gauge symmetry. 

\subsection{Lagrangian}

We work in terms of an effective heavy scalar field $\phi_v(x)$, in 
the isospin $J$ representation of $SU(2)$.    
The covariant derivative is 
$D_\mu = \partial_\mu - i g_2 W_\mu^a t^a_{J}$
and 
$W_{\mu\nu} \equiv i[D_\mu, D_\nu]/g_2 \equiv W^{a}_{\mu\nu} t^a_J$ 
is the associated field strength. 
We let $g_1$, $g_2$, $g_3 \equiv g$ denote the Standard Model 
$U(1)_Y$, $SU(2)_W$ and $SU(3)_c$ gauge coupling constants, respectively.  
A typical heavy particle momentum can be decomposed as 
\be
p^\mu = M v^\mu + k^\mu \,,
\ee
where $v^\mu$ is a velocity, $v^2=1$, and $k^\mu$ is a residual momentum.  
The basis of operators involves the perpendicular derivative, 
\be
D_{\perp}^\mu \equiv D^\mu - v^\mu v\cdot D \,.
\ee
Through $\order(1/M^3)$, 
the scalar heavy particle effective theory in the one-heavy-particle sector 
takes the form, 
\begin{align}\label{Leff}
{\cal L}_{\rm \phi} &= \phi_v^* \bigg\{ 
i v\cdot D 
- c_1 {D_\perp^2 \over 2M}
+ c_2 {D_\perp^4 \over 8M^3} 
+ {g_2 c_D} { v^\alpha [ D_\perp^\beta, W_{\alpha\beta} ] \over 8 M^2}
+ {i g_2 c_M} { \{ D_\perp^\alpha, [ D_\perp^\beta, W_{\alpha\beta} ] \} \over 16 M^3}  
\nl
&\quad
+ g_2^2 c_{A1} { W^{\alpha\beta} W_{\alpha\beta} \over 16 M^3 }
+ g_2^2 c_{A2} { v_\alpha v^\beta  W^{\mu\alpha} W_{\mu\beta} \over 16 M^3 }
+ g_2^2 c_{A3} { {\rm Tr}(W^{\alpha\beta} W_{\alpha\beta}) \over 16 M^3 }
+ g_2^2 c_{A4} {v_\alpha v^\beta {\rm Tr}(  W^{\mu\alpha} W_{\mu\beta}) \over 16 M^3 }
\nl
&\quad
+ g_2^2 c_{A1}^\prime { \epsilon^{\mu\nu\rho\sigma} W_{\mu\nu} W_{\rho\sigma} \over 16 M^3 }
+ g_2^2 c_{A2}^\prime { \epsilon^{\mu\nu\rho\sigma} v^\alpha v_\mu W_{\nu\alpha} W_{\rho\sigma} \over 16 M^3 }
+ g_2^2 c_{A3}^\prime { \epsilon^{\mu\nu\rho\sigma} {\rm Tr}(W_{\mu\nu} W_{\rho\sigma}) \over 16 M^3 }
\nl
&\quad
+ g_2^2 c_{A4}^\prime { \epsilon^{\mu\nu\rho\sigma} v^\alpha v_\mu {\rm Tr}(W_{\nu\alpha} W_{\rho\sigma}) \over 16 M^3 }
+ \dots 
\bigg\} \phi_v 
\,,
\end{align}
where we have 
employed appropriate field redefinitions to remove possible redundant operators involving 
factors of 
$v\cdot D$ acting on $\phi_v$.  
Note that the operators with coefficients $c_{A1}^\prime$ through $c_{A4}^\prime$ 
violate parity and time reversal symmetries.%
\footnote{
Additional CPT violating operators at $\order(1/M^2)$ and $\order(1/M^3)$
are constrained by reparameterization invariance to have vanishing coefficient.
} 
For the effective theory describing a fundamental heavy scalar particle, we have
$c_1=c_2=c_{A1}=1$ and $c_D=c_M=c_{A2}=c_{A3}=c_{A4}=c_{A1}^\prime=c_{A2}^\prime=c_{A3}^\prime=c_{A4}^\prime=0$ 
at tree level~\cite{Hoang:2005dk}.   
We find that the low-energy manifestation of relativistic invariance 
(``reparameterization invariance''~\cite{Luke:1992cs,Manohar:1997qy}) 
implies the exact relations,
\be
c_1 = c_2 = 1\,, 
\quad 
c_M = c_D \,.
\ee
Section~\ref{sec:fund_match} provides a nontrivial illustration of 
the latter relation. 

The complete lagrangian including Standard Model particles and interactions can be written
\be
{\cal L} = {\cal L}_\phi + {\cal L}_{\rm SM} + {\cal L}_{\phi,{\rm SM}} \,.
\ee
Here ${\cal L}_{\rm SM}$ is the usual Standard Model lagrangian, and by convention we have
included interactions with $W_\mu$ in ${\cal L}_\phi$. So far our discussion applies to a general irreducible 
$SU(2)$ representation for the heavy scalar field $\phi_v$. 
Specializing to the case of a real scalar field, 
necessarily with integer isospin, the effective theory is invariant under%
\footnote{
For a real scalar field, 
the effective theory is obtained by introducing $v_\mu$ in the field redefinition 
$\phi(x) = e^{-iMv\cdot x} \phi_v(x)/\sqrt{M} = e^{iMv\cdot x} \phi_v^*(x)/\sqrt{M} = \phi^*(x)$.   
}
\be\label{eq:vmu}
v_\mu \leftrightarrow -v_\mu \,, 
\quad
\phi_v \leftrightarrow \phi_v^* \,.
\ee
It is straightforward to verify that all interactions in ${\cal L}_\phi$ 
are invariant under this transformation. 
  
In the one-heavy-particle sector, the remaining terms involving the Higgs field $H$, 
gauge fields, and fermions are  ($\tilde{H}\equiv i\tau_2 H^*$) 
\begin{align}\label{eq:LphiSM}
&{\cal L}_{\phi,{\rm SM}} 
= 
\phi_v^* \bigg\{  c_{H} {H^\dagger H \over M} + \dots 
+ c_{Q} {t^a_J \bar{Q}_L \tau^a \slash{v} Q_L \over M^2}
+ c_{X} {i \bar{Q}_L \tau^a \gamma^\mu Q_L \{ t^a_J, D_\mu \} \over 2M^3} 
+ c_{DQ} {\bar{Q}_L \slash{v} iv\cdot D Q_L \over M^3} 
\nl
&\quad
+ c_{Du} {\bar{u}_R \slash{v} iv\cdot D u_R \over M^3} 
+ c_{Dd} {\bar{d}_R \slash{v} iv\cdot D d_R \over M^3} 
+ c_{H d} {\bar{Q}_L H d_R + h.c. \over M^3}
+ c_{H u} {\bar{Q}_L \tilde{H} u_R + h.c. \over M^3}
\nl
&\quad
+ g^2  c_{A1}^{(G)} { G^{A\,\alpha\beta} G^A_{\alpha\beta} \over 16 M^3} 
+ g^2 c_{A2}^{(G)} { v_\alpha v^\beta  G^{A\,\mu\alpha} G^A_{\mu\beta} \over 16 M^3}
+ g^2  c_{A1}^{(G)\,\prime} {\epsilon^{\mu\nu\rho\sigma} G^A_{\mu\nu} G^A_{\rho\sigma} \over 16 M^3} 
+ g^2 c_{A2}^{(G)\,\prime} { \epsilon^{\mu\nu\rho\sigma} v^\alpha v_\mu G^A_{\nu\alpha} G^A_{\rho\sigma} \over 16 M^3 }
\nl
&\quad
+ \dots 
\bigg\}\phi_v \,.
\end{align}
Terms odd under (\ref{eq:vmu}) have been omitted. Subleading terms 
containing only $H$, $\phi_v$ and their covariant derivatives
are represented by the first ellipsis in (\ref{eq:LphiSM}). 
Terms bilinear in lepton fields, and terms bilinear in the hypercharge gauge field are also 
present in ${\cal L}_{\phi,{\rm SM}}$ but have not been written explicitly. 
Repeated indices $a=1..3$ and $A=1..8$ imply a sum over gauge generators. 
Reparameterization invariance implies 
\be
c_{Q} = c_X \,.
\ee

\subsection{Sample matching calculation \label{sec:fund_match}} 

As an illustration of the construction and matching conditions 
for the heavy particle lagrangian ${\cal L}_\phi$, 
consider the case of a fundamental scalar, ignoring scalar self interactions (i.e., $\phi^4$ terms).   
For the matching of the terms containing a single gauge field, we consider 
the full theory result for the $W\phi\phi$ amputated three point function,
\vspace{2mm}
\be
\vspace{2mm}
\parbox{20mm}{\begin{fmfgraph*}(40,40)
\fmftopn{t}{3} 
\fmfbottomn{b}{3} 
\fmf{plain}{t1,v,t3}
\fmf{zigzag,label=$q$}{b2,v} 
\fmfv{label=$p,,i$}{t1}
\fmfv{label=$p^\prime,,j$}{t3}
\fmfv{label=$\mu,,a$}{b2}
\fmf{arrow_left}{t1,v}
\fmf{arrow_left}{v,t3}
\fmf{arrow_left_more}{b2,v}
\fmfblob{5mm}{v}
\end{fmfgraph*}
}
= i g_2 (p + p^\prime)^\mu F(q^2) (t_J^a)_{ji} \,,
\vspace{3mm}
\ee
where $q=p^\prime-p$, and $F(q^2)$ is a model-dependent form factor.  
Setting $p^2=p^{\prime 2}=M^2$, $v^\mu = (1,0,0,0)$, the  matching conditions 
for scalar scattering from a $\mu=0$ or $\mu=i$ gauge field read
\begin{align}\label{eq:matching}
& F(0) - F^\prime(0) {\bm{q}^2} + \dots  
 = 1 - {c_D } {\bm{q}^2 \over 8 M^2} + \dots  \,,
\nl
& {(p+p^\prime)^i} \left[ -  F(0) \left( 1 - {\bm{p}^2+ \bm{p}^{\prime 2} \over 4 M^2} \right) 
+ {F^\prime(0)} {\bm{q}^2} + \dots \right] 
\nl
&\quad 
=  {(p+p^\prime)^i} \left[ -1 + {\bm{p}^2 + \bm{p}^{\prime 2} \over 4 M^2} 
+  {c_M} {\bm{q}^2\over 8 M^2} \right]
+  {q^i} {\bm{p}^{\prime 2} - \bm{p}^2\over 8 M^2} \left( {c_D} -{c_M} \right) + \dots \,.
\end{align} 
An explicit computation of one-loop gauge boson corrections, employing 
dimensional regularization in $d=4-2\epsilon$ dimensions, yields
\be\label{eq:ff} 
F(q^2) = 1 + {g_2^2\over (4\pi)^2}{q^2\over M^2}\bigg\{ 
C_2(r)\bigg[ -{2\over 3 \epsilon_{\rm IR}} - 1 + \frac43 \log{M\over\mu} \bigg]
+ C_2(G)\bigg[ -{1\over 24 \epsilon_{\rm IR}} +\frac34 +\frac1{12}\log{M\over \mu} \bigg] \bigg\} 
+ \dots \,.
\ee
The quadratic Casimir coefficients for the isospin-$J$ and adjoint representations of $SU(2)$
are $C_2(J)=J(J+1)$ and $C_2(G)=2$. From (\ref{eq:matching}) and (\ref{eq:ff}), after 
effective theory subtractions 
the renormalized coefficients $c_D(\mu)$, $c_M(\mu)$ 
in the  $\overline{\rm MS}$ renormalization scheme are found to be
\be
c_D(\mu) = c_M(\mu) = 
{\alpha_2(\mu) \over 4\pi}  \left[ -8 J(J+1) + 12 + \left( {32 J(J+1)\over 3} +\frac43 \right)\log{M\over \mu} \right] \,. 
\ee
Matching for a general ultraviolet completion model, and for
other effective theory coefficients proceeds similarly. 

Our focus will be on the limit $M\gg m_W$, where all nontrivial matching 
conditions at the scale $\mu \sim M$ become irrelevant.%
\footnote{
In particular models with multiple mass scales, 
$1/M$ prefactors can be replaced by inverse powers of a smaller excitation energy.    
It is also of interest to investigate whether 
large anomalous dimensions could enhance the coefficients of particular subleading operators. 
}   
We leave a detailed investigation of the model-dependent form factor and
subleading $1/M$ corrections to future work. 
Presently we proceed to investigate the leading order predictions of the 
effective theory at scales $\mu \ll M$.  

\section{Integrating out the weak scale \label{sec:loweft}}

For scattering phenomena at $\sim$keV energy scales of
interest to dark matter-nucleus scattering search experiments, we should examine the
appropriate effective theory far below the electroweak scale. 
Let us begin by integrating out the degrees of freedom at the scale $m_W$.  For definiteness
we treat the top quark mass $m_t$ and the Higgs boson mass $m_h$ 
as parametrically of the same order as $m_W$. 
In following sections, we will renormalize to lower energy
scales, integrating out the remaining heavy quark degrees of freedom 
as we pass the bottom and charm quark thresholds.  The remaining hadronic matrix
elements may then be evaluated in $n_f=3$ flavor QCD to obtain 
cross section predictions. 

\subsection{Mass correction from electroweak symmetry breaking} 

We may evaluate the heavy scalar self energy to 
obtain mass corrections,    
\vspace{4mm}
\be
-i \Sigma(p) = \quad
\parbox{30mm}{\begin{fmfgraph*}(70,30)
\fmftopn{t}{3} 
\fmfbottomn{b}{3} 
\fmf{double}{b1,v1,v2,v3,b3}
\fmffreeze
\fmf{arrow_left}{b1,v1}
\fmf{zigzag,label=$W$,left}{v1,v3} 
\fmflabel{$p$}{b1}
\end{fmfgraph*}
}
+
\parbox{30mm}{\begin{fmfgraph*}(70,30)
\fmftopn{t}{3} 
\fmfbottomn{b}{3} 
\fmf{double}{b1,v1,v2,v3,b3}
\fmffreeze
\fmf{zigzag,label=$Z$,left}{v1,v3} 
\end{fmfgraph*}
}
+
\parbox{30mm}{\begin{fmfgraph*}(70,30)
\fmftopn{t}{3} 
\fmfbottomn{b}{3} 
\fmf{double}{b1,v1,v2,v3,b3}
\fmffreeze
\fmf{photon,label=$\gamma$,left}{v1,v3} 
\end{fmfgraph*}
}
+ \dots  \,. 
\ee
The shift in mass due to electroweak symmetry breaking appears as a nonvanishing value
of $\Sigma(p)$ at $v\cdot p=0$.   
We find at leading order in the $1/M$ expansion, and first order in perturbation theory,
\be\label{eq:deltaM}
\delta M 
= \alpha_2 m_W \left[ - \frac12 J^2 + \sin^2{\theta_W\over 2} J_3^2 \right] 
\,.
\ee
In particular, with $Q = J_3 + Y=J_3$ for $Y=0$, 
the mass of each charged state is lifted proportional to its squared charge 
relative to the neutral component, 
\be\label{eq:split}
M_{(Q)} - M_{(Q=0)} = {\alpha_2} Q^2 m_W \sin^2{\theta_W\over 2}  + \order(1/M)  
\approx (170 \,{\rm MeV} ) Q^2 
\,.
\ee
Subleading corrections can be similarly evaluated in the effective theory.
Since no additional operators appear at $\order(1/M^0)$, the
result (\ref{eq:split}) is model independent.%
\footnote{
The mass splitting (\ref{eq:split}) appears in limits of particular models,  
e.g. \cite{Bai:2010qg,Cheng:1998hc,Cirelli:2005uq}. 
}  

\subsection{Operator basis} 

The effective theory after electroweak symmetry breaking will include: the
heavy scalar QED theory for each of the electric charge eigenstates, with mass determined as
in (\ref{eq:deltaM});%
\footnote{
We define the pole mass to include the contributions induced by electroweak symmetry 
breaking, as opposed to introducing residual mass terms for different charge 
eigenstates~\cite{Falk:1992fm}.
}
the Standard Model lagrangian with $W,Z,h,t$
integrated out;
and interactions, 
\be
{\cal L} =   {\cal L}_{\phi_{0}} +  {\cal L}_{\rm SM}  + {\cal L}_{\phi_{0},{\rm SM}}  + \dots \,,
\ee  
where the ellipsis denotes terms containing electrically charged heavy scalars. 
For the electrically neutral scalar, 
\begin{align}
{\cal L}_{\phi_{0}} &= \phi_{v,Q=0}^{*}\bigg\{ iv\cdot \partial  
- {\partial_\perp^2\over 2M_{(Q=0)}}  
+ \order(1/m_W^3)  \bigg\} \phi_{v,Q=0} \,.
\end{align}
Note that enforcing the reality condition (\ref{eq:vmu}) implies
the vanishing of $c_D$ ($=c_M$).%

Interactions with Standard Model fields begin at order $1/m_W^3$.  
We restrict attention to quark and gluon operators (neglecting lepton and photon operators) 
and again focus on the neutral $\phi_{v,Q=0}$ component, dropping the $Q=0$ subscript in the following.  
Mixing with charged scalars will become relevant at order $1/m_W^4$ in 
nuclear scattering computations; similarly,  
we restrict attention to flavor-singlet quark bilinears, since 
matrix elements of flavor-changing bilinears are suppressed by additional 
weak coupling factors.   
Finally, we neglect operators with derivatives acting on $\phi_v$ or involving $\gamma_5$, 
since these lead to spin-dependent interactions that are suppressed for low-velocity scattering. 
The basis of operators is then 
\begin{multline}\label{eq:NRQED0}
{\cal L}_{\phi_{0},{\rm SM}} = 
{1\over m_W^3} 
\phi_v^{*} \phi_v \bigg\{ 
\sum_{q} 
\bigg[ 
c_{1q}^{(0)} O_{1q}^{(0)}   + c_{1q}^{(2)} v_\mu v_\nu O_{1q}^{(2)\mu\nu} 
\bigg]
+ c_{2}^{(0)} O_{2}^{(0)} + c_{2}^{(2)} v_\mu v_\nu O_{2}^{(2)\mu\nu} 
\bigg\}
+ \dots 
\,,
\end{multline}
where we have chosen QCD operators of definite spin,
\begin{align}\label{eq:opbasis}
O^{(0)}_{1q} &= m_q \bar{q} q \,,
\qquad & 
O^{(0)}_2 &= (G^A_{\mu\nu})^2 \,,
\nl
O^{(2)\mu\nu}_{1q} &= \bar{q}\left( \gamma^{\{\mu} iD^{\nu\}} 
- \frac{1}{d} g^{\mu\nu} i\slash{D} \right) q \,,
\qquad &
O^{(2)\mu\nu}_2 &= -G^{A \mu \lambda} G^{A \nu}_{\phantom{A \nu} \lambda} + \frac{1}{d} g^{\mu\nu} (G^A_{\alpha\beta})^2 \,.
\end{align}
Here $A^{\{\mu}B^{\nu\}} \equiv (A^\mu B^\nu + A^\nu B^\mu)/2$ denotes symmetrization. 
We employ dimensional regularization with $d=4-2\epsilon$ the spacetime dimension. 
We use the background field method for gluons in the effective theory thus 
ignoring gauge-variant operators, and assume that appropriate field redefinitions 
are employed to eliminate operators that vanish by leading order equations of motion. 
The matrix elements of the gluonic operators, 
$O^{(S)}_2$, are numerically large, representing a substantial contribution
of gluons to the energy and momentum of the nucleon. 
To account for the leading contributions from both quark and gluon operators, 
we compute the coefficients $c^{(S)}_2$ through
$\order(\alpha_s)$ and $c^{(S)}_{1q}$ through $\order(\alpha_s^0)$. 

\section{Weak scale matching \label{sec:match}} 

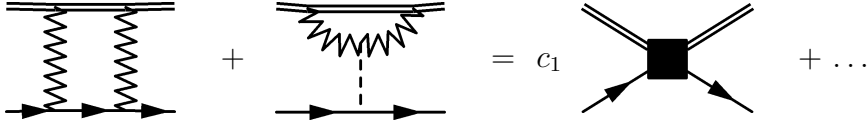
\begin{figure}
\begin{center}
\parbox{30mm}{
\begin{fmfgraph*}(80,40)
  \fmfleftn{l}{2}
  \fmfrightn{r}{2}
  \fmftopn{t}{4}
  \fmfbottomn{b}{4}
  \fmf{double}{l2,t2,t3,r2}
  \fmf{fermion}{l1,b2,b3,r1}
  \fmffreeze
  \fmf{zigzag}{t2,b2}
  \fmf{zigzag}{t3,b3}
\end{fmfgraph*}
}
+
\parbox{30mm}{
\begin{fmfgraph*}(80,40)
  \fmfleftn{l}{2}
  \fmfrightn{r}{2}
  \fmftopn{t}{5}
  \fmfbottomn{b}{5}
  \fmf{double}{l2,t2,t4,r2}
  \fmf{fermion}{l1,b3,r1}
  \fmffreeze
  \fmf{zigzag,right=0.5}{t2,v,t4}
  \fmf{dashes}{v,b3}
\end{fmfgraph*}
}
= 
$\,\,c_1\!\!$ 
\parbox{30mm}{
\begin{fmfgraph*}(80,40)
  \fmfleftn{l}{2}
  \fmfrightn{r}{2}
  \fmftopn{t}{4}
  \fmfbottomn{b}{4}
  \fmf{double}{l2,v,r2}
  \fmf{fermion}{l1,v,r1}
  \fmffreeze
  \fmfv{decor.shape=square}{v}
\end{fmfgraph*} 
} + \dots
\caption{\label{fig:quarkmatching}
Matching condition for quark operators.  
Double lines denote heavy scalars, zigzag lines denote 
$W$ bosons, dashed lines denote Higgs bosons, 
single lines with arrows denote quarks, and the solid square denotes an effective theory 
vertex.   Diagrams with crossed $W$ lines are not displayed. 
}
\end{center}
\end{figure}

The matching conditions for quark operators in the $n_f=5$ flavor theory 
at renormalization 
scale $\mu=\mu_t  \sim m_t \sim m_W \sim m_h$ are obtained from 
the diagrams in Fig.~(\ref{fig:quarkmatching}):
\begin{align}\label{eq:quarkmatch}
c_{1U}^{(0)}(\mu_t) &= {\cal C} \left[ - {1\over x_h^2}  \right] \,,
\quad
& c_{1D}^{(0)}(\mu_t) &= {\cal C} \left[ - {1\over x_h^2 } - |V_{tD}|^2 {x_t\over 4(1+x_t)^3}  
 \right] \,,
\nl
c_{1U}^{(2)}(\mu_t) &= {\cal C} \left[ \frac23 \right] \,,  
\quad 
& c_{1D}^{(2)}(\mu_t) &= {\cal C} \left[ \frac23  
- |V_{tD}|^2 {  x_t ( 3 + 6 x_t + 2 x_t^2)\over 3(1+x_t)^3} 
 \right] \,,  
\end{align}
where subscript $U$ denotes $u$ or $c$ and subscript $D$ denotes $d$, $s$ or $b$.    
Here ${\cal C}=  [\pi \alpha_2^2(\mu_t)][J(J+1)/2]$, $x_h\equiv m_h/m_W$ and $x_t\equiv m_t/m_W$.  
We ignore corrections of order $m_q/m_W$ for $q=u,d,s,c,b$, and 
have used CKM unitarity to simplify the results. 

\begin{figure}
\begin{center}
\parbox{30mm}{
\begin{fmfgraph*}(80,80)
  \fmfleftn{l}{3}
  \fmfrightn{r}{3}
  \fmftopn{t}{4}
  \fmfbottomn{b}{4}
  \fmf{double}{l3,t2,t3,r3}
  \fmf{phantom,tag=3}{l1,r1}
  \fmf{phantom,tag=4}{l3,r3}
  \fmffreeze
  \fmf{phantom,tension=3}{l2,v1}
  \fmf{phantom,tension=3}{r2,v2}
  \fmf{phantom,left,tag=1}{v1,v2}
  \fmf{phantom,left,tag=2}{v2,v1}
  \fmfposition
  \fmffreeze
  \fmfipath{p[]}
  \fmfiset{p1}{vpath1(__v1,__v2)}
  \fmfiset{p2}{vpath2(__v2,__v1)}
  \fmfiset{p3}{vpath3(__l1,__r1)}
  \fmfiset{p4}{vpath4(__l3,__r3)}
  \fmfi{fermion}{subpath (length(p1)/2, 3length(p1)/2) of (p1 & p2)}
  \fmfi{fermion}{subpath (length(p1)/2, 3length(p1)/2) of (p2 & p1)}
  \fmfi{gluon}{point length(p2)/3 of p2 -- point 2length(p3)/3 of p3 }
  \fmfi{gluon}{point 2length(p2)/3 of p2 -- point length(p3)/3 of p3 }
  \fmfi{zigzag}{point length(p1)/6 of p1 -- point length(p4)/5 of p4 } 
  \fmfi{zigzag}{point 5length(p1)/6 of p1 -- point 4length(p4)/5 of p4 } 
\end{fmfgraph*}
}
+
\parbox{30mm}{
\begin{fmfgraph*}(80,80)
  \fmfleftn{l}{3}
  \fmfrightn{r}{3}
  \fmftopn{t}{4}
  \fmfbottomn{b}{4}
  \fmf{double}{l3,t2,t3,r3}
  \fmf{phantom,tag=3}{l1,r1}
  \fmf{phantom,tag=4}{l3,r3}
  \fmffreeze
  \fmf{phantom,tension=3}{l2,v1}
  \fmf{phantom,tension=3}{r2,v2}
  \fmf{phantom,left,tag=1}{v1,v2}
  \fmf{phantom,left,tag=2}{v2,v1}
  \fmfposition
  \fmffreeze
  \fmfipath{p[]}
  \fmfiset{p1}{vpath1(__v1,__v2)}
  \fmfiset{p2}{vpath2(__v2,__v1)}
  \fmfiset{p3}{vpath3(__l1,__r1)}
  \fmfiset{p4}{vpath4(__l3,__r3)}
  \fmfi{fermion}{subpath (length(p1)/2, 3length(p1)/2) of (p1 & p2)}
  \fmfi{fermion}{subpath (length(p1)/2, 3length(p1)/2) of (p2 & p1)}
  \fmfi{gluon}{point 2length(p2)/5 of p1 -- point length(p3)/3 of p3 }
  \fmfi{gluon}{point 3length(p2)/5 of p1 -- point 2length(p3)/3 of p3 }
  \fmfi{zigzag}{point length(p1)/6 of p1 -- point length(p4)/5 of p4 } 
  \fmfi{zigzag}{point 5length(p1)/6 of p1 -- point 4length(p4)/5 of p4 } 
\end{fmfgraph*}
}
+ 
\parbox{30mm}{
\begin{fmfgraph*}(80,80)
  \fmfleftn{l}{3}
  \fmfrightn{r}{3}
  \fmftopn{t}{4}
  \fmfbottomn{b}{4}
  \fmf{double}{l3,t2,t3,r3}
  \fmf{phantom,tag=3}{l1,r1}
  \fmf{phantom,tag=4}{l3,r3}
  \fmffreeze
  \fmf{phantom,tension=3}{l2,v1}
  \fmf{phantom,tension=3}{r2,v2}
  \fmf{phantom,left,tag=1}{v1,v2}
  \fmf{phantom,left,tag=2}{v2,v1}
  \fmfposition
  \fmffreeze
  \fmfipath{p[]}
  \fmfiset{p1}{vpath1(__v1,__v2)}
  \fmfiset{p2}{vpath2(__v2,__v1)}
  \fmfiset{p3}{vpath3(__l1,__r1)}
  \fmfiset{p4}{vpath4(__l3,__r3)}
  \fmfi{fermion}{subpath (length(p1)/2, 3length(p1)/2) of (p1 & p2)}
  \fmfi{fermion}{subpath (length(p1)/2, 3length(p1)/2) of (p2 & p1)}
  \fmfi{gluon}{point 2length(p2)/3 of p2 -- point length(p3)/3 of p3 }
  \fmfi{gluon}{point length(p1)/2 of p1 -- point 2length(p3)/3 of p3 }
  \fmfi{zigzag}{point length(p1)/6 of p1 -- point length(p4)/5 of p4 } 
  \fmfi{zigzag}{point 5length(p1)/6 of p1 -- point 4length(p4)/5 of p4 } 
\end{fmfgraph*}
}
+
\parbox{30mm}{
\begin{fmfgraph*}(80,80)
  \fmfleftn{l}{3}
  \fmfrightn{r}{3}
  \fmftopn{t}{5}
  \fmfbottomn{b}{5}
  \fmf{double}{l3,t2,t4,r3}
  \fmf{phantom,tag=3}{l1,r1}
  \fmf{phantom,tag=4}{l3,r3}
  \fmffreeze
  \fmf{phantom,tension=2}{t3,v1}
  \fmf{phantom,tension=5}{b3,v2}
  \fmf{phantom,left,tag=1}{v1,v2}
  \fmf{phantom,left,tag=2}{v2,v1}
  \fmfposition
  \fmffreeze
  \fmfipath{p[]}
  \fmfiset{p1}{vpath1(__v1,__v2)}
  \fmfiset{p2}{vpath2(__v2,__v1)}
  \fmfiset{p3}{vpath3(__l1,__r1)}
  \fmfiset{p4}{vpath4(__l3,__r3)}
  \fmfi{fermion}{p1}
  \fmfi{fermion}{p2}
  \fmfi{gluon}{point length(p2)/6 of p2 -- point length(p3)/3 of p3 }
  \fmfi{gluon}{point 5length(p1)/6 of p1 -- point 2length(p3)/3 of p3 }
  \fmf{zigzag,right=0.5}{t2,xx,t4}
  \fmf{dashes}{xx,v1}
\end{fmfgraph*}
}

\vspace{5mm}
= 
$\,\,c_2\!\!$ 
\parbox{30mm}{
\begin{fmfgraph*}(80,40)
  \fmfleftn{l}{2}
  \fmfrightn{r}{2}
  \fmftopn{t}{4}
  \fmfbottomn{b}{4}
  \fmf{phantom}{b1,b2,b3,b4}
\fmffreeze
  \fmf{double}{l2,v,r2}
  \fmf{gluon, tension=0.5}{b2,v,b3}
  \fmffreeze
  \fmfv{decor.shape=square}{v}
\end{fmfgraph*} 
} 
+
$\,\,c_1\! \Bigg[$ 
\parbox{30mm}{
\begin{fmfgraph*}(80,60)
  \fmfleftn{l}{3}
  \fmfrightn{r}{3}
  \fmftopn{t}{3}
  \fmfbottomn{b}{3}
  \fmf{phantom}{l3,t2,r3}
  \fmf{phantom}{l1,b2,r1}
  \fmf{phantom,tag=3}{l1,r1}
  \fmf{phantom,tension=5}{t2,b2}
  \fmf{phantom,tension=10}{t2,v1}
  \fmf{phantom,tension=6}{b2,v2}
  \fmf{plain,left,tag=1}{v1,v2}
  \fmf{plain,left,tag=2}{v2,v1}
  \fmf{phantom,tag=4}{l3,v1}  
  \fmf{phantom,tag=5}{v1,r3}  
  \fmfposition
  \fmffreeze
   \fmfipath{p[]}
   \fmfiset{p1}{vpath1(__v1,__v2)}
   \fmfiset{p2}{vpath2(__v2,__v1)}
   \fmfiset{p3}{vpath3(__l1,__r1)}
   \fmfiset{p4}{vpath4(__l3,__v1)}
   \fmfiset{p5}{vpath5(__v1,__r3)}
   \fmfi{fermion}{subpath (0,length(p1)) of p1}
   \fmfi{fermion}{subpath (0,length(p2)) of p2}
   \fmfi{gluon}{point 2length(p1)/3 of p1 -- point 2length(p3)/3 of p3 }
   \fmfi{gluon}{point length(p2)/3 of p2 -- point length(p3)/3 of p3 }
   \fmfi{double}{subpath (0,length(p4)) of p4}
   \fmfi{double}{subpath (0,length(p5)) of p5}
  \fmfv{decor.shape=square}{v1}
\end{fmfgraph*}
}
+
\parbox{30mm}{
\begin{fmfgraph*}(80,60)
  \fmfleftn{l}{3}
  \fmfrightn{r}{3}
  \fmftopn{t}{3}
  \fmfbottomn{b}{3}
  \fmf{phantom}{l3,t2,r3}
  \fmf{phantom}{l1,b2,r1}
  \fmf{phantom,tag=3}{l1,r1}
  \fmf{phantom,tension=5}{t2,b2}
  \fmf{phantom,tension=10}{t2,v1}
  \fmf{phantom,tension=6}{b2,v2}
  \fmf{plain,left,tag=1}{v1,v2}
  \fmf{plain,left,tag=2}{v2,v1}
  \fmf{phantom,tag=4}{l3,v1}  
  \fmf{phantom,tag=5}{v1,r3}  
  \fmfposition
  \fmffreeze
   \fmfipath{p[]}
   \fmfiset{p1}{vpath1(__v1,__v2)}
   \fmfiset{p2}{vpath2(__v2,__v1)}
   \fmfiset{p3}{vpath3(__l1,__r1)}
   \fmfiset{p4}{vpath4(__l3,__v1)}
   \fmfiset{p5}{vpath5(__v1,__r3)}
   \fmfi{fermion}{subpath (0,length(p1)) of p1}
   \fmfi{fermion}{subpath (0,length(p2)) of p2}
   \fmfi{gluon}{point length(p2)/3 of p2 -- point length(p3)/3 of p3 }
   \fmfi{gluon}{point 0 of p1 -- point 2length(p3)/3 of p3 }
   \fmfi{double}{subpath (0,length(p4)) of p4}
   \fmfi{double}{subpath (0,length(p5)) of p5}
  \fmfv{decor.shape=square}{v1}
\end{fmfgraph*}
}
$\Bigg]$
+ \dots
\caption{\label{fig:gluonmatching}
Matching condition onto gluon operators.  The notation is as in 
Fig.~\ref{fig:quarkmatching}. 
}
\end{center}
\end{figure}
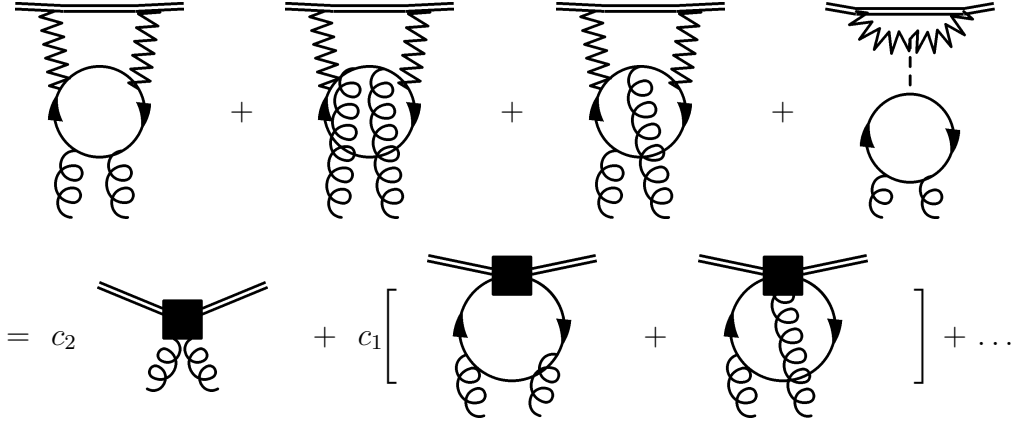

Matching conditions onto gluon operators are from the diagrams of Fig.~(\ref{fig:gluonmatching}): 
\begin{align}
c_{2}^{(0)}(\mu_t) &={\cal C} {\alpha_s(\mu_t) \over 4\pi} 
\left[  
{1 \over 3 x_h^2}   
+ {3 + 4 x_t + 2 x_t^2 \over 6(1+x_t)^2 }
\right] \,,
\nl
c_{2}^{(2)}(\mu_t) &= {\cal C} {\alpha_s(\mu_t) \over 4\pi} 
\bigg[   
 - {32\over 9} \log{\mu_t\over m_W} - 4 
- {4 (2+ 3x_t)\over 9(1+x_t)^3}\log{\mu_t \over m_W(1+x_t)}
\nl
&\quad
-{4 ( 12 x_t^5 - 36 x_t^4 + 36 x_t^3 - 12 x_t^2 + 3 x_t - 2)\over 9 (x_t-1)^3}\log{x_t\over 1+x_t}
- {8 x_t ( -3 + 7 x_t^2) \over 9(x_t^2-1)^3} \log 2
\nl
&\quad
- { 48 x_t^6 + 24 x_t^5 - 104 x_t^4 - 35 x_t^3 + 20 x_t^2 + 13 x_t + 18 \over 9(x_t^2-1)^2 (1+x_t)}
\bigg] \,. 
\end{align}
There is no dependence of $c_2^{(0)}$ or $c_2^{(2)}$ 
on CKM matrix elements in the limit of vanishing $d,s,b$ quark masses. 
The renormalized coefficients are computed in the $\overline{\rm MS}$ scheme.
We have employed Fock-Schwinger ($x\cdot A=0$) gauge~\cite{Novikov:1983gd} to compute the
full-theory amplitudes for gluonic operators in Fig.~\ref{fig:gluonmatching}. 
The effective theory subtractions 
are efficiently performed in a scheme with massless light quarks, using
dimensional regularization as infrared regulator. We have verified that the  
same results are obtained using finite masses and taking the limit $m_q/m_W \to 0$.  
Details of this computation will be presented elsewhere. 

\section{RG evolution to hadronic scales \label{sec:RG} }

To account for perturbative corrections involving large logarithms, 
e.g. $\alpha_s(\mu_0) \log{m_t/\mu_0}$, 
we employ renormalization group evolution to sum leading logarithms to 
all orders.  
 
\subsection{Anomalous dimensions} 

The spin $S=0$ and spin $S=2$ operators mix amongst themselves, with
\be
{d\over d\log{\mu}} O_i^{(S)} = - \sum_j \gamma^{(S)}_{ij} O_j \,,
\ee
where $\gamma^{(S)}_{ij}$ are $(n_f+1)\times (n_f+1)$ anomalous dimension matrices. 
The leading terms are 
\begin{align}\label{eq:anom}
\hat{\gamma}^{(0)} 
&=  \left( \begin{array}{ccc|c} 
0 & & & 0 \\
& \ddots & & \vdots \\
&& 0 & 0 \\
\hline
-2\gamma_m^\prime & \cdots & -2\gamma_m^\prime & (\beta/g)^\prime 
\end{array} 
\right)
= {\alpha_s \over 4\pi} \left( \begin{array}{ccc|c} 
0 & & & 0 \\
& \ddots & & \vdots \\
&& 0 & 0 \\
\hline
32 & \cdots & 32 & -2 \beta_0 
\end{array} 
\right)  + \dots \,,
\nl
\hat{\gamma}^{(2)} 
&= 
{\alpha_s \over 4\pi} \left( \begin{array}{ccc|c}
{64\over 9} & & & -\frac43 \\
& \ddots & & \vdots \\
&& {64\over 9} & -\frac43 \\
\hline
-{64\over 9} & \cdots & -{64\over 9} & {4n_f\over 3} 
\end{array} 
\right) + \dots \,,
\end{align}
where $\beta = {d g/ d \log\mu} \approx - \beta_0 {\alpha_s/ 4\pi}$, 
$\gamma_m = {d \log m_q / d\log \mu} \approx -8 {\alpha_s / 4\pi}$, 
$\gamma_m^\prime \equiv g \partial \gamma_m /\partial g$, 
$(\beta/g)^\prime \equiv g \partial (\beta/g) /\partial g$,  
and $\beta_0 = 11 - \frac23 n_f$.  
It is straightforward to include subleading terms for 
$\hat{\gamma}^{(0)}$~\cite{Tarrach:1981bi,Grinstein:1988wz} 
and $\hat{\gamma}^{(2)}$~\cite{Vogt:2004mw,Moch:2004pa}.

\subsection{Integrating out heavy quarks} 

At the scale $\mu = \mu_b \sim m_b$, we match onto an $n_f=4$ theory containing $u,d,s,c$ quarks. 
The matching equations are
\begin{align} 
& c_{2}^{(0)}(\mu_b) = \tilde{c}_{2}^{(0)}(\mu_b)\left(1+  {4 \tilde{a}\over 3}  \log \frac{m_b}{\mu_b} \right) 
-  {\tilde{a}\over 3} \tilde{c}_{1b}^{(0)}(\mu_b)  
\left[ 1+  \tilde{a} \left( 11 + \frac43 \log \frac{m_b}{\mu_b} \right) \right] + \mathcal{O}(\tilde{a}^3),
\nl 
& c_{1q}^{(0)}(\mu_b) = \tilde{c}_{1q}^{(0)}(\mu_b) + \mathcal{O}(\tilde{a}^2),
\nl
& c_{2}^{(2)}(\mu_b) = \tilde{c}_2^{(2)}(\mu_b) -\frac{4\tilde{a}}3 \log{m_b\over \mu_b}
\tilde{c}_{1b}^{(2)}(\mu_b)  + \mathcal{O}(\tilde{a}^2), 
\nl
& c_{1q}^{(2)}(\mu_b) = \tilde{c}_{1q}^{(2)}(\mu_b)  + \mathcal{O}(\tilde{a}), 
\end{align} 
where $q=u,d,s,c$ and $\tilde{a}=\alpha_s(\mu_b,n_f=5)/4\pi$. 
Quantities without (with) tilde refer to 
the $n_f=4$ ($n_f=5$) theory. 
The scheme dependence for heavy quark masses enters at higher order. 
For definiteness we use pole masses for $m_b$ and $m_c$, with values taken from 
\cite{Martin:2009iq}.
Following our power counting scheme, we consider one less order of 
$\alpha_s$ in the matching for $c_{1q}^{(S)}$ relative to $c_{2}^{(S)}$. 
For later use in the numerical analysis, 
we have included NLO QCD corrections in the spin-$0$ matching 
\cite{Ovrut:1981bg,Inami:1982xx}.
Similar to above, we evolve coefficients in the $n_f=4$ theory to the scale $\mu= \mu_c \sim m_c$. 
Finally, we match onto $n_f=3$ and evolve to a low scale $\mu_0 \sim 1\,{\rm GeV}$ independent of heavy quark masses.

\section{Matrix elements and cross section \label{sec:xs}} 

Having expressed the lagrangian in terms of operators renormalized at 
the scale $\mu_0 \sim 1\,{\rm GeV}$, we require hadronic matrix elements 
evaluated at this scale.

\subsection{Hadronic inputs \label{sec:inputs}}

Let us define the zero-momentum matrix elements of renormalized operators%
\footnote{
We use nonrelativistic normalization for nucleon states, 
$\langle N(\bm{p})|N(\bm{p}^\prime)\rangle = (2\pi)^3\delta^3(\bm{p}-\bm{p}^\prime)$.
}
\begin{align} \label{eq:hadme}
\langle N| O_{1q}^{(0)}| N \rangle &\equiv m_N f_{q,N}^{(0)} \,,
\qquad 
{- 9\alpha_s(\mu)\over 8\pi} \langle N | O_2^{(0)}(\mu) | N \rangle \equiv m_N f_{G,N}^{(0)}(\mu) \,,
\nl
\langle N | O_{1q}^{(2)\mu\nu}(\mu) | N \rangle 
&\equiv {1\over m_N} \left(k^\mu k^\nu - {g^{\mu\nu} \over 4} m_N^2 \right) f_{q,N}^{(2)}(\mu) \,,
\nl
\langle N | O_2^{(2)\mu\nu}(\mu) | N \rangle 
&\equiv {1\over m_N} \left(k^\mu k^\nu - {g^{\mu\nu} \over 4} m_N^2 \right) f_{G,N}^{(2)}(\mu) \,,
\end{align}
where $m_N$ is the nucleon mass.  Matrix elements refer to a definite (but arbitrary) spin state of the nucleon. 

\subsubsection{Spin zero}

\begin{table}[t]
\begin{center}
\begin{tabular}{c|c|c}
Parameter & Value & Reference\\
\hline
$|V_{td}|$ & $\sim 0$ & -\\
$|V_{ts}|$ & $\sim 0$ & -\\ 
$|V_{tb}|$ & $\sim 1$ & -\\
${m_u/ m_d}$ & $0.49(13)$ & \cite{Nakamura:2010zzi} \\
${m_s/ m_d}$ & $19.5(2.5)$ & \cite{Nakamura:2010zzi} \\
$\Sigma_{\pi N}^{\rm lat}$ & $0.047(9)\,{\rm GeV}$ & \cite{Young:2009zb} \\
$\Sigma_s^{\rm lat}$ & $0.050(8)\,{\rm GeV}$ & \cite{Giedt:2009mr} \\
$\Sigma_{\pi N}$ & $0.064(7)\,{\rm GeV}$ &\cite{Pavan:2001wz} \\
$\Sigma_0$ & $0.036(7)\,{\rm GeV}$ &\cite{Borasoy:1996bx}\\
$m_W$ & $80.4 \,{\rm GeV}$ & \cite{Nakamura:2010zzi}\\
$m_t$ & 172 GeV & \cite{Martin:2009iq}\\
$m_b$ & 4.75 GeV & \cite{Martin:2009iq}\\
$m_c$ & 1.4 GeV & \cite{Martin:2009iq}\\
$m_N$ & 0.94 GeV &-\\
$\alpha_s(m_Z)$ & 0.118 & \cite{Nakamura:2010zzi}\\
$\alpha_2(m_Z)$ & 0.0338 & \cite{Nakamura:2010zzi}\\
\end{tabular} 
\end{center}
\caption{\label{tab:inputs}
Inputs to the numerical analysis.
} 
\end{table}

We recall that the spin-0 operator matrix elements are not independent, 
being linked by the relation~\cite{Shifman:1978zn} 
\begin{align}\label{eq:trace}
m_N
= (1-\gamma_m) \sum_q \langle N | m_q \bar{q} q | N \rangle 
+ {\beta\over 2g } \langle N | (G^a_{\mu\nu})^2 | N \rangle  \,,
\end{align} 
derived from the trace of the QCD energy-momentum tensor. 
Here $N=p$ or $n$. 
Neglecting $\gamma_m$, $\order(\alpha_s^2)$ contributions to $\beta(g)$, 
and power corrections in the above formula, the definitions (\ref{eq:hadme})  
ensure that
$f_{G,N}^{(0)}(\mu) \approx 1 - \sum_{q=u,d,s} f_{q,N}^{(0)}$.
Corrections 
arising from (\ref{eq:trace}) 
are included in the numerical analysis.

For quark operators, define the scale-independent quantities, 
\begin{align} 
\Sigma_{\pi N} = 
{m_u + m_d \over 2} \langle p | (\bar{u} u + \bar{d} d) | p \rangle 
\,, \quad 
\Sigma_0 =
{m_u + m_d \over 2} \langle p | (\bar{u} u + \bar{d} d - 2\bar{s} s ) | p \rangle \,. 
\end{align} 
In the numerical analysis, we will neglect the small contributions proportional to $|V_{td}|^2$ 
and $|V_{ts}|^2$, so that $c_{1u}^{(0)} = c_{1d}^{(0)}$.   
Neglecting also the small contribution~\cite{Gasser:1982ap}  
$(m_d-m_u) \langle p | (\bar{u} u - \bar{d} d) | p\rangle \sim 2\,{\rm MeV}$, 
and using approximate isospin symmetry, 
we then require, for $N=p$ or $n$,
\begin{align}
m_N ( f_{u,N}^{(0)} + f_{d,N}^{(0)} ) &\approx  \Sigma_{\pi N}  \,, \quad
m_N f_{s,N}^{(0)} =  {m_s  \over {m_u + m_d} } ( \Sigma_{\pi N} - \Sigma_0) = \Sigma_s\,. 
\end{align}
We consider ``traditional'' values $\Sigma_{\pi N} = 64\pm 7 \,{\rm MeV}$~\cite{Pavan:2001wz} and 
$\Sigma_0 = 36 \pm 7\,{\rm MeV}$~\cite{Borasoy:1996bx},
but investigate also the lattice determinations, 
$\Sigma_{\pi N}^{\rm lat} = 47 \pm 9 \, {\rm MeV}$~\cite{Young:2009zb} 
and $\Sigma_s^{\rm lat} = 50 \pm 8 \,{\rm MeV}$~\cite{Giedt:2009mr}.%
\footnote{The latter quantity arises from a naive averaging of 
$\Sigma_s  = 31 \pm 15 \, {\rm MeV}$~\cite{Young:2009zb} 
and $\Sigma_s = 59 \pm 10\,{\rm MeV}$~\cite{Toussaint:2009pz}.  See also \cite{Takeda:2010cw,Ohki:2008ff,Durr:2011mp}. 
} 
We adopt PDG values~\cite{Nakamura:2010zzi} for light-quark mass ratios. 
A summary of numerical inputs is presented in Table~\ref{tab:inputs}.

\subsubsection{Spin two}

\begin{table}[t]
\begin{center}
\begin{tabular}{c|cccc}
 $\mu \text{(GeV)}$ & $f_{u,p}^{(2)}(\mu)$ & $f_{d,p}^{(2)}(\mu)$ & $f_{s,p}^{(2)}(\mu)$ & $f_{G,p}^{(2)}(\mu)$ 
\\
\hline 
1.0 & 0.404(6) & 0.217(4) & 0.024(3) & 0.36(1)
\\
1.2 & 0.383(6) & 0.208(4) & 0.027(2) & 0.38(1)
\\
1.4 & 0.370(5) & 0.202(4) & 0.030(2) & 0.40(1)
\end{tabular} 
\end{center}
\caption{\label{tab:momfraction}
Operator coefficients derived from MSTW PDF analysis \cite{Martin:2009iq} at different
values of $\mu$. 
} 
\end{table}

The matrix elements of spin-two operators can be identified as 
\be
f_{q,p}^{(2)}(\mu) = \int_0^1 dx \, x [ q(x,\mu) + {\bar{q}}(x,\mu) ] \,,
\ee
where $q(x,\mu)$ and $\bar{q}(x,\mu)$ are parton distribution functions 
evaluated at scale $\mu$. 
Neglecting power corrections, the sum of spin two operators in (\ref{eq:opbasis}) is the
traceless part of the QCD energy momentum tensor, hence independent of scale we have
$f_{G,p}^{(2)}(\mu) \approx 1 - \sum_{q=u,d,s} f_{q,p}^{(2)}(\mu)$.
Using approximate isospin symmetry we set
\be
f_{u,n}^{(2)} = f_{d,p}^{(2)} \,, \quad
f_{d,n}^{(2)} = f_{u,p}^{(2)} \,, \quad
f_{s,n}^{(2)} = f_{s,p}^{(2)} \,.
\ee
Table~\ref{tab:momfraction} lists coefficient values for
renormalization scales $\mu=1\,{\rm GeV}$, $\mu=1.2\,{\rm GeV}$ 
and $\mu=1.4\,{\rm GeV}$ determined from the parameterization and analysis of \cite{Martin:2009iq}.

\subsection{Cross section}

The low-velocity, spin-independent, cross section for 
WIMP scattering on a nucleus of mass number $A$ and charge $Z$ may be written
\be\label{eq:fullxs}
\sigma_{A,Z} = {m_r^2\over \pi} \left|  Z {\cal M}_p + (A-Z) {\cal M}_n \right|^2 
\approx {m_r^2 A^2 \over \pi} |{\cal M}_p|^2 
\,,
\ee
where ${\cal M}_p$ and ${\cal M}_n$ are the matrix elements for scattering on a proton
or neutron respectively%
\footnote{ Explicitly, ${\cal M}_N = m_W^{-3} \langle N| \left(\sum_{q=u,d,s} 
\left[ c_{1q}^{(0)} O^{(0)}_{1q} + c_{1q}^{(2)} v_\mu v_\nu O^{(2) \mu \nu}_{1q}  \right] 
+ c^{(0)}_{2}  O^{(0)}_{2}  + c^{(2)}_{2}v_\mu v_\nu  O^{(2)\mu \nu}_{2}   \right) | N \rangle$
.}, 
and $m_r = M m_{\cal N} /(M + m_{\cal N})$ 
denotes the reduced mass of the dark-matter nucleus system.  
As described in Section~\ref{sec:inputs}, ${\cal M}_n \approx {\cal M}_p$ up to 
corrections from numerically small CKM factors and isospin violation in nucleon matrix
elements.  In the $M \gg m_{\cal N}$ limit, the cross section scales as $A^4$. 
At finite velocity, a nuclear form factor modifies this behavior~\cite{Jungman:1995df}. 

\begin{figure}
\begin{center}
\includegraphics[width=12cm]{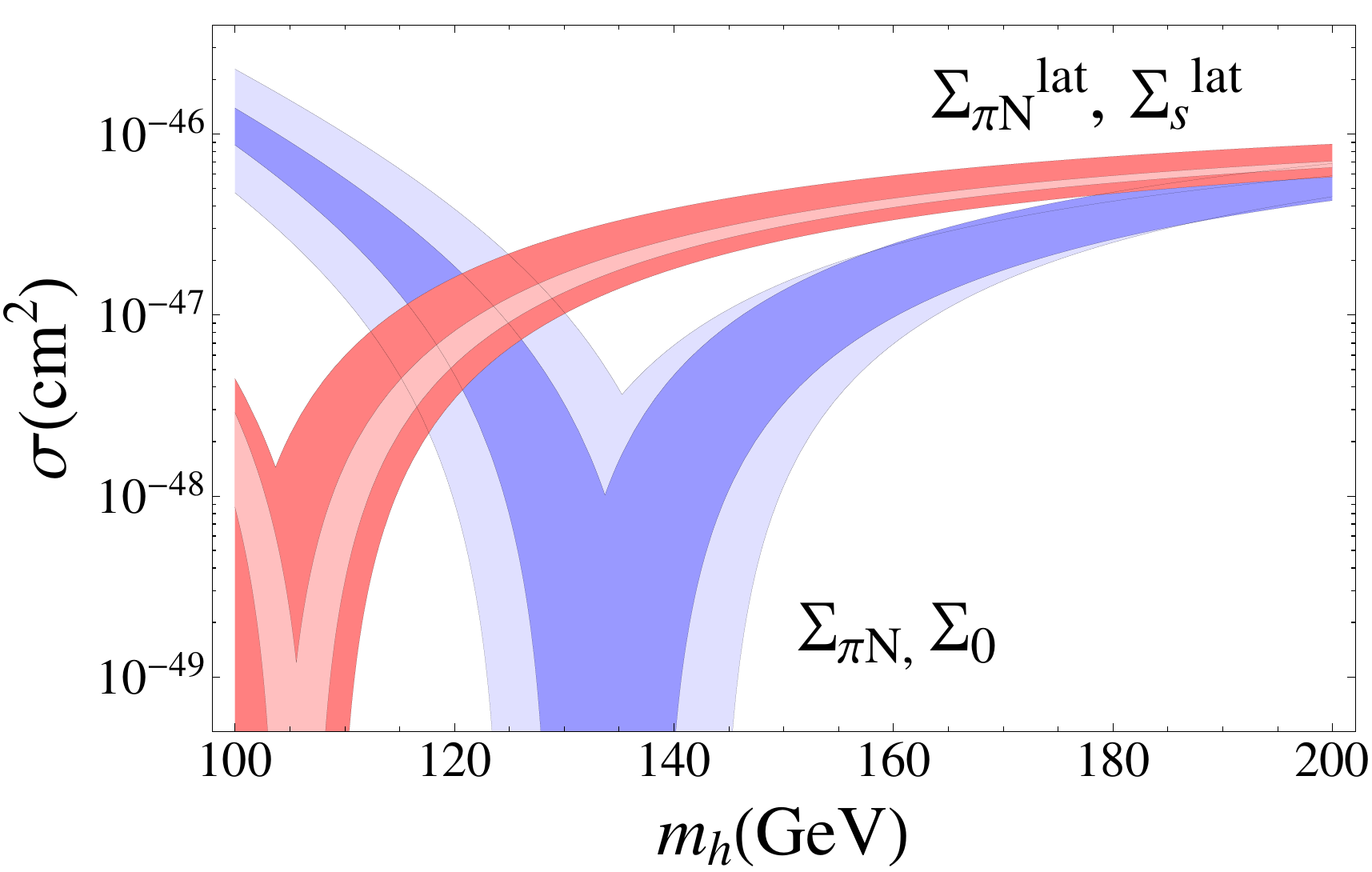}
\caption{\label{fig:xs}
Cross section for low-velocity scattering on a nucleon for a heavy real scalar in the isospin $J=1$ representation of $SU(2)$.   
The dark shaded region represents the $1\sigma$ 
uncertainty from perturbative QCD, estimated by varying factorization scales.  
The light shaded region represents the $1\sigma$ uncertainty from hadronic inputs.  
}
\end{center}
\end{figure}

As a numerical benchmark, let us compute the cross section
for low-momentum scattering on a nucleon for a heavy real scalar in the isospin representation $J=1$. 
Figure~\ref{fig:xs} displays the result, as a function of the unknown 
Higgs boson mass.   
Using Table~\ref{tab:inputs}, 
we consider separately the ``traditional'' inputs $\Sigma_{\pi N}$ and $\Sigma_0$, 
as well as recent lattice determinations of $\Sigma_{\pi N}^{\rm lat}$ and $\Sigma_s^{\rm lat}$.   
For each case, separate bands represent the 
uncertainty due to neglected perturbative QCD corrections, 
and due to the hadronic inputs. 
We estimate the impact of 
higher order perturbative QCD corrections by varying matching scales 
$m_W^2/2 \le \mu_t^2 \le 2m_t^2$, $m_b^2/2 \le \mu_b^2 \le 2 m_b^2$, 
$m_c^2/2 \le \mu_c^2 \le 2 m_c^2$, $1.0 \,{\rm GeV} \le \mu_0 \le 1.4 \,{\rm GeV}$, 
adding the errors in quadrature.  

The renormalization group running and heavy quark matching for
spin-$2$ operators are evaluated at LO.%
\footnote{ Up to power corrections and subleading
$\order(\alpha_s)$ corrections, our evaluation is equivalent to an
evaluation in either the $n_f=4$ or  $n_f=5$ flavors theories, taking
the $c$- and $b$-quark momentum fractions of the proton as input.  We
have verified that these results,  with the matrix elements taken from
\cite{Martin:2009iq}, are within  our error budget.    
}  
For spin-$0$ operators, we find a large residual uncertainty at LO 
from $\mu_0$, $\mu_c$ and $\mu_b$ scale variation. 
The RG running from  $\mu_c$ to $\mu_0$ from (\ref{eq:anom}) 
is thus evaluated with NNNLO corrections, including contributions 
to $\beta/g$ through  $\order(\alpha_s^4)$ and $\gamma_m$ through
$\order(\alpha_s^4)$.  
Accordingly, the spin-$0$ gluonic matrix element from (\ref{eq:trace})
is also evaluated at NNNLO, including contributions to $\beta/g$ through
$\order(\alpha_s^4)$ and $\gamma_m$ through $\order(\alpha_s^3)$. 
The residual $\mu_0$ scale variation is insignificant compared to other 
uncertainties.
We perform the RG running and heavy quark matching 
from $\mu_t$ to $\mu_c$ at NLO. 
Hadronic input uncertainties from each source in Table~\ref{tab:inputs} 
and Table~\ref{tab:momfraction} are added in quadrature. 
We have ignored power corrections appearing at relative order 
$\alpha_s(m_c)\Lambda_{\rm QCD}^2/m_c^2$; typical numerical prefactors appearing 
in the coefficients of the corresponding power-suppressed operators~\cite{Shifman:1978zn} 
suggest that these effects are small. 

\begin{figure}
\begin{center}
\includegraphics[width=120mm]{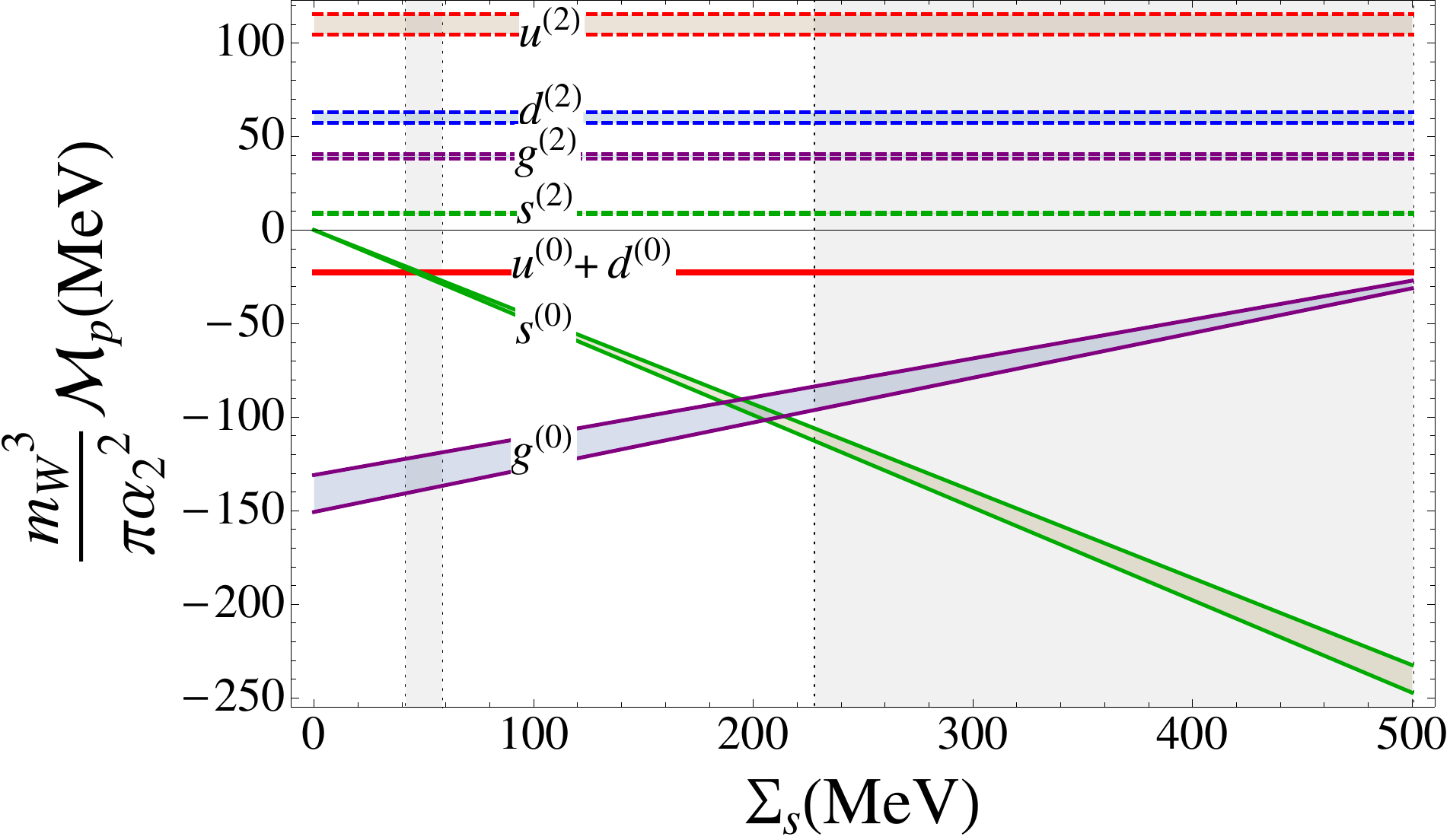}
\caption{\label{fig:bands}
Breakdown of contributions to the 
matrix element ${\cal M}_p$ using the representative values $m_h=120\,{\rm GeV}$ and $\Sigma_{\pi N}^{\rm lat} = 47(9) \,{\rm MeV}$.    The labels $u^{(S)}$, $d^{(S)}$, $s^{(S)}$ and $g^{(S)}$ refer to spin-S up, down, strange and gluon operator contributions, respectively. The thickness represents the $1\sigma$ uncertainty from perturbative QCD. The left-hand
vertical band corresponds to the lattice value $\Sigma_s^{\rm lat} = 50(8) \,{\rm MeV}$ 
and the right-hand vertical band corresponds to the range $\Sigma_s = 366(142) \,{\rm MeV}$  
deduced from $\Sigma_{\pi N}$ and $\Sigma_{0}$ in Table~\ref{tab:inputs}.
}
\end{center}
\end{figure}

Due to a partial cancellation between spin-0 and spin-2 matrix elements, 
the total cross section and the fractional error
depend sensitively on subleading perturbative corrections and 
on the Higgs mass parameter $m_h$. 
We find
\be
\sigma_p(m_h=120\,{\rm GeV}) = 0.7 \pm 0.1 {}^{+0.9}_{-0.3} \times 10^{-47}{\rm cm}^2  
\,,
\quad 
\sigma_p(m_h=140\,{\rm GeV}) = 2.4 \pm 0.2 {}^{+1.5}_{-0.6} \times 10^{-47}{\rm cm}^2 \,,
\ee
where the first error is from hadronic inputs, assuming $\Sigma_s^{\rm lat}$ and 
$\Sigma_{\pi N}^{\rm lat}$ from Table~\ref{tab:inputs}, and the second error 
represents the effect of neglected higher order perturbative QCD corrections.  
For the illustrative value $m_h=120\,{\rm GeV}$, and as a function of
the scalar strange-quark matrix element $\Sigma_s$,
we display the separate contributions 
of each of the quark and gluon operators in Fig.~\ref{fig:bands}.  

\section{Summary \label{sec:summary}}

We have presented the effective theory 
for heavy, weakly interacting dark matter candidates charged under electroweak $SU(2)$. 
Having determined the general form of the effective lagrangian (\ref{Leff}) through
$1/M^3$, we demonstrated matching conditions for subleading operators 
in a simple model.  
Using the effective theory, we demonstrated universality of the 
mass splitting induced by electroweak symmetry breaking, and of the cross section
for scattering on nuclear matter.  
Subleading terms in the $1/M$ expansion can be studied systematically 
using (\ref{Leff}).    

Our focus has been on the case of an isotriplet real scalar~\cite{Bai:2010qg}.   For this case, 
relic abundance estimates~\cite{Cirelli:2005uq} 
indicate that $M \lesssim {\rm few}\, {\rm TeV}$ 
in order to not overclose the universe.   This mass range, combined with
the universal cross section, provides a target for future search experiments. 

We have presented a complete matching at first nonvanishing order in $\alpha_s$, 
and at leading order in small ratios $m_W/M$, $m_b/m_W$ and $\Lambda_{\rm QCD}/m_c$.  
We performed renormalization group improvement 
to sum leading logarithms to all orders.   
The residual dependence on the high matching 
scale $\mu_t \sim m_t \sim m_W $ represents uncertainty due to uncalculated higher-order
perturbative corrections.  
Assuming the hadronic input $\Sigma_s^{\rm lat}$ from Table~\ref{tab:inputs}, this scale variation 
is the largest remaining uncertainty on the cross section; its reduction would require higher loop order
calculations.    

Our high-scale matching results for
quark operators (\ref{eq:quarkmatch}) 
and spin-zero gluon operators agree with $m_W/M\to 0$ results
presented  by Hisano et al.~\cite{Hisano:2011cs}, under the 
identification $\mu_t=\mu_b=\mu_c$, i.e., a one-step matching onto the $n_f=3$ theory.%
\footnote{
To make the comparison to the scattering amplitude for a heavy 
Majorana fermion with $\chi = \chi^c$, 
we use $\chi = \sqrt{2} e^{-imv\cdot x} (h_v + H_v) = \sqrt{2} e^{im v\cdot x} (h_v^c + H_v^c)$,
where $h_v$ and $H_v$ are spinor fields with $(1-\slash{v})h_v = (1+\slash{v})H_v = 0$.   
} 
This approach neglects large logarithms appearing in coefficient functions. 
The effective theory analysis provides a systematically improvable method to resum 
large logarithms, and to estimate both perturbative and hadronic-input uncertainties.
Our results for matching onto spin-two gluon operators are new. 

With obvious modifications, our results can be extended to higher $SU(2)$ representations (or 
$SU(2)$ singlets), fermionic WIMPs, and models with additional low-energy field content 
beyond the Standard Model. 
The treatment of QCD corrections presented here can be applied to
compute scattering amplitudes in related models, e.g. models involving
coupling to hypercharge, or supersymmetric models~\cite{Jungman:1995df,Drees:1992rr,Ellis:2008hf}. 

\vskip 0.2in
\noindent
{\bf Acknowledgements}
\vskip 0.1in
\noindent
We acknowledge discussions with Y. Bai, G. Paz and J. Zupan. 
Work supported by NSF Grant 0855039. 

\vskip 0.2 in
\noindent
{\it Note added.} 
While this paper was in writing, the preprint \cite{Kopp:2011gg}
appeared, mentioning the invariance (\ref{eq:vmu}) for Majorana fermions.

\end{fmffile} 


\begin{thebibliography}{99} 

\bibitem{Bai:2010qg}
  Y.~Bai and R.~J.~Hill,
  Phys.\ Rev.\  D {\bf 82}, 111701 (2010)
  [arXiv:1005.0008 [hep-ph]].

\bibitem{DMmodels} 
  Other models containing dark matter $SU(2)$ multiplets have been 
  studied in:
  J.~Bagnasco, M.~Dine, S.~D.~Thomas,
  Phys.\ Lett.\  {\bf B320}, 99-104 (1994).
  [hep-ph/9310290].
%
  M.~Cirelli, N.~Fornengo and A.~Strumia,
  Nucl.\ Phys.\  B {\bf 753}, 178 (2006)
  [arXiv:hep-ph/0512090].
%
  T.~Hur, D.~W.~Jung, P.~Ko and J.~Y.~Lee,
  Phys.\ Lett.\  B {\bf 696}, 262 (2011)
  [arXiv:0709.1218 [hep-ph]].
%
  R.~Essig,
  Phys.\ Rev.\  D {\bf 78}, 015004 (2008)
  [arXiv:0710.1668 [hep-ph]].
%
  C.~Kilic, T.~Okui and R.~Sundrum,
  JHEP {\bf 1002}, 018 (2010)
  [arXiv:0906.0577 [hep-ph]].
%
  M.~T.~Frandsen, F.~Sannino,
  Phys.\ Rev.\  {\bf D81}, 097704 (2010).
  [arXiv:0911.1570 [hep-ph]].
%
  T.~Cohen, J.~Kearney, A.~Pierce, D.~Tucker-Smith,
  [arXiv:1109.2604 [hep-ph]].
%

\bibitem{Isgur:1989vq}
  N.~Isgur, M.~B.~Wise,
  Phys.\ Lett.\  {\bf B232}, 113 (1989).
%
  W.~E.~Caswell, G.~P.~Lepage,
  Phys.\ Lett.\  {\bf B167}, 437 (1986).
%
  E.~Eichten, B.~R.~Hill,
  Phys.\ Lett.\  {\bf B234}, 511 (1990).
    
\bibitem{Hoang:2005dk}
  Tree level matching for a heavy colored scalar was considered in:
  A.~H.~Hoang and P.~Ruiz-Femenia,
  Phys.\ Rev.\  D {\bf 73}, 014015 (2006)
  [arXiv:hep-ph/0511102].

\bibitem{Luke:1992cs}
  M.~E.~Luke, A.~V.~Manohar,
  Phys.\ Lett.\  {\bf B286}, 348-354 (1992).
  [hep-ph/9205228].

\bibitem{Manohar:1997qy}
  A.~V.~Manohar,
  Phys.\ Rev.\  {\bf D56}, 230-237 (1997).
  [hep-ph/9701294].

\bibitem{Cheng:1998hc}
  H.~-C.~Cheng, B.~A.~Dobrescu, K.~T.~Matchev,
  Nucl.\ Phys.\  {\bf B543}, 47-72 (1999).
  [hep-ph/9811316].

\bibitem{Cirelli:2005uq}
  M.~Cirelli, N.~Fornengo, A.~Strumia,
  Nucl.\ Phys.\  {\bf B753}, 178-194 (2006).
  [hep-ph/0512090].

\bibitem{Falk:1992fm}
  A.~F.~Falk, M.~Neubert, M.~E.~Luke,
  Nucl.\ Phys.\  {\bf B388}, 363-375 (1992).
  [hep-ph/9204229].

\bibitem{Novikov:1983gd}
  V.~A.~Novikov, M.~A.~Shifman, A.~I.~Vainshtein, V.~I.~Zakharov,
  Fortsch.\ Phys.\  {\bf 32}, 585 (1984).

\bibitem{Tarrach:1981bi}
  R.~Tarrach,
  Nucl.\ Phys.\  {\bf B196}, 45 (1982).

\bibitem{Grinstein:1988wz}
  B.~Grinstein, L.~Randall,
  Phys.\ Lett.\  {\bf B217}, 335 (1989).

\bibitem{Vogt:2004mw}
  A.~Vogt, S.~Moch, J.~A.~M.~Vermaseren,
  Nucl.\ Phys.\  {\bf B691}, 129-181 (2004).
  [hep-ph/0404111].

\bibitem{Moch:2004pa}
  S.~Moch, J.~A.~M.~Vermaseren, A.~Vogt,
  Nucl.\ Phys.\  {\bf B688}, 101-134 (2004).
  [hep-ph/0403192].
  
\bibitem{Martin:2009iq}
  A.~D.~Martin, W.~J.~Stirling, R.~S.~Thorne and G.~Watt,
  Eur.\ Phys.\ J.\  C {\bf 63}, 189 (2009)
  [arXiv:0901.0002 [hep-ph]].

\bibitem{Ovrut:1981bg}
  B.~A.~Ovrut, H.~J.~Schnitzer,
  Phys.\ Lett.\  {\bf B110}, 139 (1982).
  
\bibitem{Inami:1982xx}
  T.~Inami, T.~Kubota, Y.~Okada,
  Z.\ Phys.\ C\  {\bf 18}, 69-80 (1983).

\bibitem{Shifman:1978zn}
  M.~A.~Shifman, A.~I.~Vainshtein, V.~I.~Zakharov,
  Phys.\ Lett.\  {\bf B78}, 443 (1978).

\bibitem{Gasser:1982ap}
  J.~Gasser, H.~Leutwyler,
  Phys.\ Rept.\  {\bf 87}, 77-169 (1982).

\bibitem{Nakamura:2010zzi}
  K.~Nakamura {\it et al.} [ Particle Data Group Collaboration ],
  J.\ Phys.\ G {\bf G37}, 075021 (2010).

\bibitem{Young:2009zb}
  R.~D.~Young and A.~W.~Thomas,
  Phys.\ Rev.\  D {\bf 81}, 014503 (2010)
  [arXiv:0901.3310 [hep-lat]].

\bibitem{Giedt:2009mr}
  J.~Giedt, A.~W.~Thomas and R.~D.~Young,
  Phys.\ Rev.\ Lett.\  {\bf 103}, 201802 (2009)
  [arXiv:0907.4177 [hep-ph]].
  
\bibitem{Pavan:2001wz}
  M.~M.~Pavan, I.~I.~Strakovsky, R.~L.~Workman, R.~A.~Arndt,
  PiN Newslett.\  {\bf 16}, 110-115 (2002).
  [hep-ph/0111066].

\bibitem{Borasoy:1996bx}
  B.~Borasoy, U.~-G.~Meissner,
  Annals Phys.\  {\bf 254}, 192-232 (1997).
  [hep-ph/9607432].

\bibitem{Toussaint:2009pz}
  D.~Toussaint and W.~Freeman  [MILC Collaboration],
  Phys.\ Rev.\ Lett.\  {\bf 103}, 122002 (2009)
  [arXiv:0905.2432 [hep-lat]].

\bibitem{Takeda:2010cw}
  K.~Takeda {\it et al.} [ JLQCD Collaboration ],
  Phys.\ Rev.\  {\bf D83}, 114506 (2011).
  [arXiv:1011.1964 [hep-lat]].

\bibitem{Ohki:2008ff}
  H.~Ohki {\it et al.},
  Phys.\ Rev.\  D {\bf 78}, 054502 (2008)
  [arXiv:0806.4744 [hep-lat]].

\bibitem{Durr:2011mp}
  S.~Durr, Z.~Fodor, T.~Hemmert, C.~Hoelbling, J.~Frison, S.~D.~Katz, S.~Krieg, T.~Kurth {\it et al.},
  ``Sigma term and strangeness content of octet baryons,''
    arXiv:1109.4265 [hep-lat].

\bibitem{Jungman:1995df}
  G.~Jungman, M.~Kamionkowski and K.~Griest,
  Phys.\ Rept.\  {\bf 267}, 195 (1996)
  [arXiv:hep-ph/9506380].

\bibitem{Hisano:2011cs}
  J.~Hisano, K.~Ishiwata, N.~Nagata, T.~Takesako,
  JHEP {\bf 1107}, 005 (2011).
  [arXiv:1104.0228 [hep-ph]].
 %
  J.~Hisano, K.~Ishiwata, N.~Nagata,
  Phys.\ Rev.\  {\bf D82}, 115007 (2010).
  [arXiv:1007.2601 [hep-ph]].

\bibitem{Drees:1992rr}
  M.~Drees, M.~M.~Nojiri,
  Phys.\ Rev.\  {\bf D47}, 4226-4232 (1993).
  [hep-ph/9210272].
  M.~Drees, M.~Nojiri,
  Phys.\ Rev.\  {\bf D48}, 3483-3501 (1993).
  [hep-ph/9307208].
  
\bibitem{Ellis:2008hf}
  J.~R.~Ellis, K.~A.~Olive and C.~Savage,
  Phys.\ Rev.\  D {\bf 77}, 065026 (2008)
  [arXiv:0801.3656 [hep-ph]].

\bibitem{Kopp:2011gg}
  K.~Kopp, T.~Okui,
  [arXiv:1108.2702 [hep-ph]].

\end{thebibliography}
\end{document}